\documentclass[arXiv]{actapoly}

\usepackage{color}
\usepackage{xspace}
\usepackage{graphicx}
\usepackage{natbib}
\usepackage{amsmath}
\usepackage{microtype}

\newcommand{\gro}{GRO~J1008--57\xspace}

\newcommand{\tb}{$\bullet$ \footnotesize}

\begin{document}

\title{Simultaneous fits in ISIS on the example of \gro}

\institution{remeis}{Remeis-Observatory \& ECAP, Universit\"at Erlangen-N\"urnberg, Sternwartstr. 7, 96049 Bamberg, Germany}
\institution{mit}{MIT Kavli Institute for Astrophysics, Cambridge, MA 02139, USA}
\institution{cresst}{CRESST, Center for Space Science and Technology, UMBC, Baltimore, MD 21250, USA}
\institution{gsfc}{NASA Goddard Space Flight Center, Greenbelt, MD 20771, USA}
\institution{ucsd}{Center for Astronomy and Space Sciences, University of California, San Diego, La Jolla, CA 92093, USA}
\institution{iaat}{Institut f\"ur Astronomie und Astrophysik, Universit\"at T\"ubingen, Sand 1, 72076 T\"ubingen, Germany}
\institution{isdc}{ISDC Data Center for Astrophysics, Chemin d'\'Ecogia 16, 1290 Versoix, Switzerland}

\correspondingauthor{M.~K\"uhnel}{remeis}{matthias.kuehnel@sternwarte.uni-erlangen.de}
\author{S.~M\"uller}{remeis}
\author{I.~Kreykenbohm}{remeis}
\author{F.-W.~Schwarm}{remeis}
\author{C.~Gro{\ss}berger}{remeis}
\author{T.~Dauser}{remeis}
\author{M.~A.~Nowak}{mit}
\author{K.~Pottschmidt}{gsfc,cresst}
\author{C.~Ferrigno}{isdc}
\author{R.~E.~Rothschild}{ucsd}
\author{D.~Klochkov}{iaat}
\author{R.~Staubert}{iaat}
\author{J.~Wilms}{remeis}

\begin{abstract}
Parallel computing and steadily increasing computation speed have led to a new
tool for analyzing multiple datasets and datatypes: fitting several datasets
simultaneously. With this technique, physically connected parameters of
individual data can be treated as a single parameter by implementing this
connection into the fit directly. We discuss the terminology, implementation,
and possible issues of simultaneous fits based on the X-ray data analysis
tool \texttt{Interactive Spectral Interpretation System} (ISIS). While all data
modeling tools in X-ray astronomy allow in principle fitting data from multiple
data sets individually, the syntax used in these tools is not often well suited
for this task.

Applying simultaneous fits to the transient X-ray binary \gro, we find
that the spectral shape is only dependent on X-ray flux. We determine
time independent parameters such as, e.g., the folding energy $E_\text{fold}$,
with unprecedented precision.
\end{abstract}


\maketitle

\section{Motivation}\label{sec:motivation}

Most data analysis in X-ray astronomy concentrate on describing single datasets
or on characterizing samples with results of fits of individual datasets. Once a
good description of an example dataset is found, the analysis of comparable
datasets follows.
Finally, the results of all those individual analyses are compared and
interpreted.

For instance, a particular parameter is found to depend on other parameters.
Instead of going back to the data analysis and fitting this dependency
directly to enhance the parameter precision or break degeneracies (feasible
through reduced degrees of freedom), the dependency is then analyzed on its own. In
another way, the former analysis is indeed repeated but with this
parameter fixed according to the discovered dependency. Furthermore, if
parameters cannot be constrained well, it is common to keep those parameters
fixed to a certain standard value.

Thus, one cannot gain any physical information from this fixed parameter
and, more importantly, systematical effects might arise. The reason for not
following sophisticated ways is usually a lack of computation power. Implementing
parameter correlations or dependencies would require one to analyze all data
at the same time. However, since computer power has increased and parallel
computation using several computers is possible, this situation has changed
today. In other words, fitting data simultaneously has become feasible even when
large numbers of datasets (e.g., 50-100 pointings at a single source) are to be
considered.

\begin{table}
\caption{Advantages and disadvantages of fitting several datasets simultaneously.}
\centering
\begin{tabular}{p{.42\linewidth}p{.46\linewidth}}
\hline
advantages & disadvantages \\
\hline\hline
\tb fixed parameters can be determined correctly & \tb increased runtime of fits and uncertainty calculations  \\
\tb complicated parameter correlations can be implemented and tested & \tb large memory is needed $\rightarrow$ multi-CPU calculations required \\
\tb combination of different types of data is possible & \tb statistical weights of datasets have to be choosen \\
\tb parameter degeneracies can be broken &  \tb careful handling of fit-parameters required \\
\tb reduced number of degrees of freedom & \\
\hline
\end{tabular}
\label{tab:procontra}
\end{table}

In Section~\ref{sec:implementation} we introduce an implementation of
simultaneous data analysis into the \texttt{Interactive Spectral Interpretation System}
 \citep[ISIS,][]{houck2000a}, which has been ``designed to facilitate the
interpretation and analysis of high resolution X-ray spectra''\footnote{http://space.mit.edu/CXC/ISIS/}.
In Section~\ref{sec:applications}, we present ideas for possible applications
of simultaneous data analysis and further demonstrate the power of this method on the
example of the transient X-ray binary \gro in Section~\ref{sec:gro1008}.
Finally, we discuss questions and issues, which arise by comparing
advantages and disadvantages of simultaneous fits (Table~\ref{tab:procontra}).

\section{Implementation into ISIS}\label{sec:implementation}

ISIS \cite{houck2000a} was developed to fit X-ray spectra, but it can also be used
to analyze nearly all kinds of data due to its strong customization capability
\cite{noble2008a} compared to, e.g., XSPEC \cite{schafer1991a,arnaud1996a}. For instance,
user-defined fit-functions, as well as complex correlations between data and models, can
be implemented. However, functions to handle these correlations for a large number
of parameters and datasets in an easy way are not yet available.

Before we describe the technical realization of simultaneous data analysis
in ISIS, we introduce new notations used by the implemented functions.

\subsection{Terminology}

The parameters of a model which is fitted to data either act on
all datasets loaded into ISIS, or on an individual dataset. By defining
parameters for each dataset and tying them to each other, parameters can be
linked to multiple datasets similar to the approach chosen, e.g., in XSPEC.

We call multiple datasets, which should be fitted with the same set of
parameters, a \textit{datagroup}. The corresponding parameters are called
\textit{group parameters}. \textit{Global parameters} denote
parameters which act on all datagroups. 

\begin{figure}
\centering
\includegraphics[width=.9\linewidth]{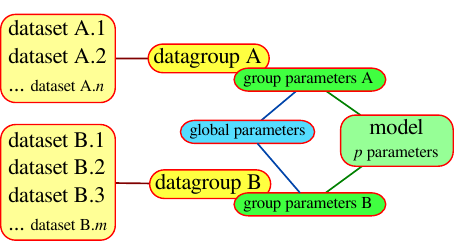}
\caption{Terminology of simultaneous fits in ISIS. There are $n$
 and $m$ simultaneous datasets, forming the datagroup A and B, respectively.
Both datasets have their own group parameters, resulting from a model with
$p$ parameters. Some of the group parameters are the same for both
datagroups and are called global parameters.}
\label{fig:terminology}
\end{figure}

Figure~\ref{fig:terminology} illustrates these definitions. In this example,
a dataset requires a model with $p$ parameters. There are simultaneous data
from $n$ detectors available which can be described by the same parameters.
These datasets define the datagroup A with $p$ free parameters. Another group
of data was recorded by $m$ detectors. These datasets define an individual
datagroup B with, again, $p$ free parameters. During the analysis of
both groups, however, it turns out that a specific parameter seems to be equal
for both data groups within the uncertainties. As a result, the two
individual values for this parameter are tied to each other, resulting in a
global parameter. That reduces the number of free parameters by one and the
remaining group parameters can be constrained better.

\subsection{Data- and analysis functions}

Since simultaneous fits can have large numbers of fit parameters connected
by a complicated logic, we provide a collection of all functions necessary
to initialize and perform simultaneous fits in ISIS\footnote{these functions
are available as part of the \texttt{isisscripts}, a collection of useful
functions, which can be downloaded at
http://www.sternwarte.uni-erlangen.de/git.public/?p=isisscripts}.
The initialization of a simultaneous fit is performed via
\begin{verbatim}
 simfit = simultaneous_fit();
\end{verbatim}
where \verb|simultaneous_fit| returns a structure (\texttt{Struct\_Type}),
which has to be assigned to a variable, here \verb|simfit|. The structure
contains several
functions and fields to handle simultaneous fits. The documentation of each
function is available using the \verb|help|-qualifier. Some important functions
are described in the following.

\begin{verbatim}
 simfit.add_data(filenames);
\end{verbatim}
This defines a datagroup and loads the spectra given by \verb|filenames|, which must
be an array of strings. The function also allows other data than spectra to be
loaded or defined.

\begin{verbatim}
 simfit.fit_fun(model);
\end{verbatim}
The string \verb|model| defines the fit-function to be used for all
datasets. Here, the placeholder \% can be used instead of a component
instance. In this case individual group parameters are applied to each
datagroup automatically.

\begin{verbatim}
 simfit.set_par_fun(parameter, function);
\end{verbatim}
This is probably one of the most useful functions. Like the ISIS intrinsic
function, the value of the \verb|parameter| is determined by the given
\verb|function|. The \%-placeholder can be used within the string
\verb|parameter| to apply the function to the corresponding parameter of
each data group. However, the function may contain other parameters or even a
single parameter name as well. In the latter case, if the function is also
applied to all datagroups (using the \%), the single parameter is treated as
global parameter from now on.\\

Because a simultaneous fit results in a large number of parameters, a single
call to a fit-routine (\verb|fit_counts|) will take a long time. In the
example of the previous section, the final model fitted to the data consists
of $(n+m)\times p$ parameters, where only $2 p - 1$ are free. To reduce the
runtime of a fit, three fit-routines are implemented within the simultaneous-
fit-structure.

\begin{verbatim}
 simfit.fit_groups(groupID);
\end{verbatim}

Instead of perfoming a $\chi^2$-minimization of all parameters and datasets,
this function loops over all datagroups and fits only the associated parameters
(group parameters). If a group is specified by the optional \verb|groupID|,
then only the group parameters of this particular group are fitted.

\begin{verbatim}
 simfit.fit_global();
\end{verbatim}
Instead of fitting the group parameters, this function fits the global
parameters only. Since all defined data groups have to be taken into
account, the fit usually takes longer than fitting the group parameters. \\

\subsection{Uncertainty calculations}\label{sec:uncertainties}

As already mentioned, the runtime of simultaneous fits is
increased compared to fitting a single dataset only. Thus, uncertainty
calculations of parameters, where a certain parameter's range has to be found
corresponding to a change in $\chi^2$, will be affected dramatically by the
high runtime. Furthermore, it is necessary to distinguish between group- and
global parameters. We recommend to compute the uncertainty intervals for each
parameter on a different machine by, e.g., using \citep{noble2006a} or
\texttt{mpi\_fit\_pars} and the \texttt{SLmpi}
module\footnote{http://www.sternwarte.uni-erlangen.de/\\git.public/?p=slmpi.git}.
We compared the runtime of a parallel uncertainty calculation in ISIS with a
serial approach in XSPEC. Estimating the uncertainties of 10 parameters in
parallel (i.e., on 10 cores) is faster by more than a factor of 3 (21\,ks vs. 60\,ks).
Additionally, the calculations in ISIS resulted in a better $\chi^2_\mathrm{red}$
because the parameter ranges being scanned are larger in the parallel approach.

Group parameters depend on a single datagroup only. As a consequence, all other
datagroups and therefore all other group parameters can be ignored during the
uncertainty calculation. Unfortunately, that is not the case for global
parameters. During the analysis of \gro (see Section~\ref{sec:gro1008}), the
uncertainties of the global parameters have been calculated by revealing the
$\chi^2$-landscape of each global parameter by individual fits. Afterwards,
every landscape has been interpolated to find the $\Delta \chi^2$-value of
interest (e.g., $\Delta \chi^2 = 2.71$ corresponding to the 90\%-confidence
level).
In this way the runtime of an uncertainty calculation of a single global
parameter could be reduced significantly. Note that depending on the model
and amount of data, such a computation can take up to several days.

\section{Applications}\label{sec:applications}

There are various applications of simultaneous fits and data analysis. Besides
determining specific parameters which seem to be constant over time by all
available data, more physical questions can be tackled. For instance, if a
physical property of the object of interest results in multiple
observables:

\begin{itemize}
 \item the geometry of the accretion column in accreting neutron star X-ray
       binaries affects the line shape of cyclotron resonant scattering features
       (CRSF) \citep{schoenherr2007a} as well as the pulse profile shape (Falkner et
       al., in prep.).
\end{itemize}

Furthermore, instead of deriving physical properties from parameters after fits
have been performed, these properties can be directly fitted to the data by
implementing the dependency on the model parameters:
\begin{itemize}
 \item the components in radio maps of jets in active galactic
       nuclei move with certain velocities. Assuming a constant velocity of
       the jet components, the velocity itself could be a global fit parameter
       \cite{grossberger}.
 \item in the sub-critical accretion regime of neutron stars, the spectrum is
       believed to harden with increasing luminosity \citep{becker2012a}.
       Any possible dependency between power-law shape and luminosity could be
       fitted simultaneously with multiple spectra.
\end{itemize}

\section{The Example \gro}\label{sec:gro1008}

As an example of a successful simultaneous fit we briefly summarize the results
of our analysis of \mbox{\gro} using almost all available X-ray spectra and
-lightcurves. This transient high-mass X-ray binary consists of a neutron star
orbiting a Be-type optical companion. For further details of the system as well
as for the results of the analysis see \cite{kuehnel2013a} and references therein.
% 
% \begin{figure}
% \centering
% \includegraphics[width=\linewidth]{joined_confmaps}
% \caption{Confidence contours of the $\chi^2$-landscape of the orbital
% period and time of periastron passage. Shown are 68\% (solid), 90\% (dashed) and
% 99\% (dotted) confidence levels. The fit including the outburst times (green)
% results in better uncertainties as compared to one without those times (red).}
% \label{fig:contour}
% \end{figure}

Since sources are only visible for a small fraction of their full orbit, it is
challenging to obtain the orbital parameters of transient X-ray binaries by
analyzing, e.g., pulse arrival times \citep{deeter1981a,boynton1986a}.
Thus, an observed shift in the orbital phase with respect to initial
orbital parameters can be fitted with either a different orbital period or
time of periastron passage. This leads to a parameter degeneracy,
which can be visualized by a contour map of the $\chi^2$-landscape of these
parameters. %The red contour lines in Fig.~\ref{fig:contour} show
%that both parameters are degenerated statistically (i.e., the ellipse
%is tilted).
The resulting contour map shows that both parameters are degenerated
statistically (i.e., the ellipsoidal contour lines are tilted).

The outburst times of the source are, however, clearly connected to the
periastron passage. Performing a simultaneous fit of the pulse arrival times
and the outburst times reduces the parameter degeneracy and results in much
better constrained parameters (about a factor of 2-3) as seen by %the green contour lines in
%Fig.~\ref{fig:contour}.
recalculated contour map.

\begin{figure}
\centering
\includegraphics[width=\linewidth]{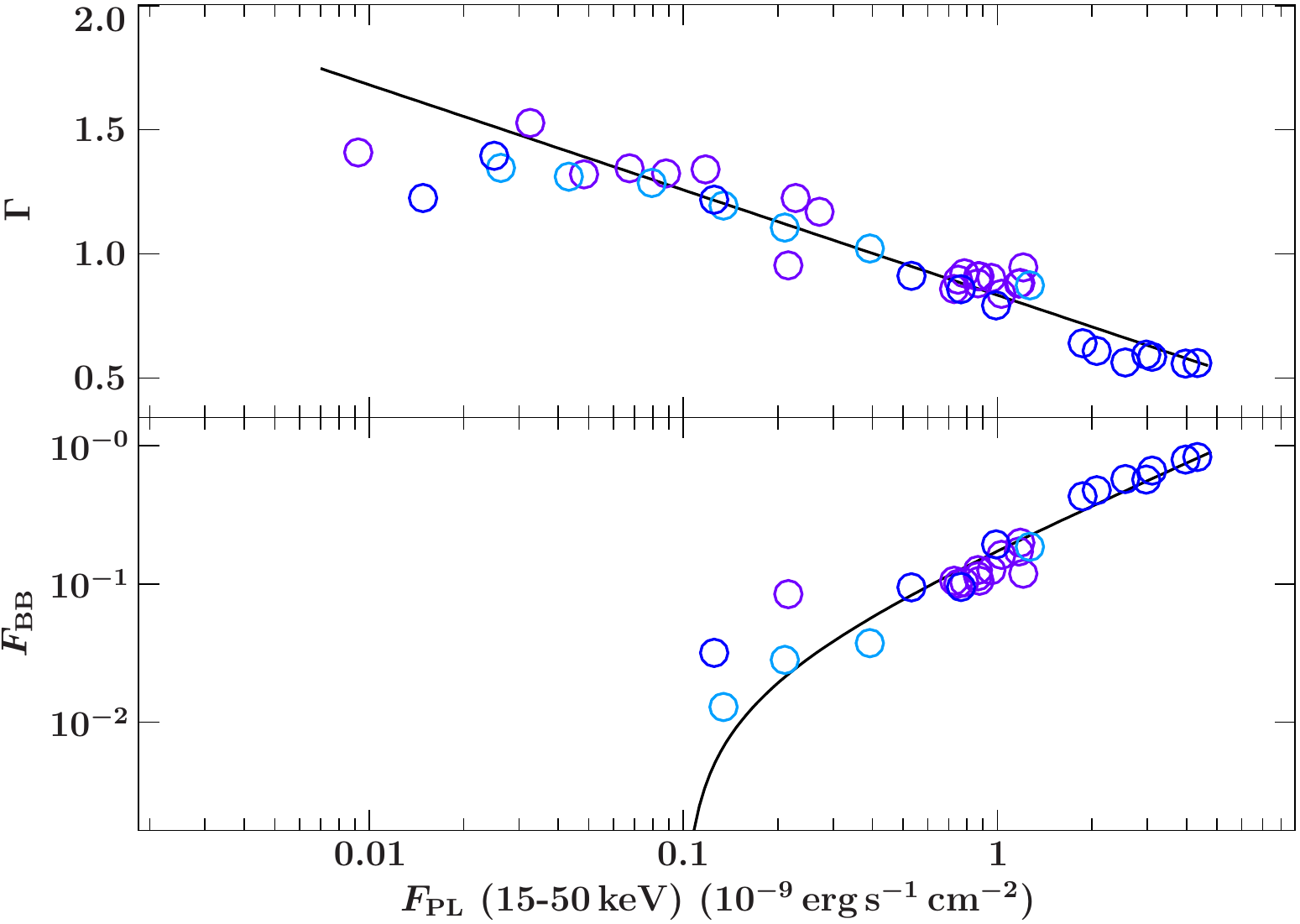}
\caption{A fit (black lines) of the power-law index $\Gamma$ and black body
flux $F_\text{BB}$ as functions of power-law flux $F_\text{PL}$. The different
colors correspond to different outbursts.}
\label{fig:paramfit}
\end{figure}

Initial fits of the spectra of three outbursts of \gro in 2005, 2007 and 2011
with an absorbed cutoff power-law and an additional black body component showed
that the folding energy $E_\text{fold}$, as well as the black body temperature
$kT$, are independent of time within uncertainties. In particular, it seems
that they do not change between different outbursts, i.e., these parameters are
constant properties of the source.

Thus, those parameters are set as global parameters using
\verb|simfit.set_par_fun| and their values are determined by all available
data. In addition, further parameters can be treated as global ones
\citep[see][for more details]{kuehnel2013a}. Finally, each observation
is described by 3 group parameters only ($\approx$ degrees of freedom for each
datagroup, the global parameters contribute marginally), which are the power-law
flux $F_\text{PL}$, the black body flux $F_\text{BB}$, and the photon index
$\Gamma$. The latter two strongly correlate with $F_\text{PL}$, but show
no dependency on the outburst time or -shape. This correlation can be fitted
to describe the spectrum of \gro by only one parameter: the power-law flux
$F_\text{PL}$. The fit is shown in Fig.~\ref{fig:paramfit} and its values
are given in Section 4.2 of \citep{kuehnel2013a}.

As already mentioned in Section~\ref{sec:uncertainties}, the runtime of
uncertainty calculations of the global parameters is increased dramatically.
In the case of this analysis, the $\chi^2$-landscape produced by taking all
68 spectra into account was interpolated to estimate the uncertainties. The
calculation of a single global parameter took $\sim\!7$\,days on 100~CPUs
(16320\,CPUh).

\section{Outlook}\label{sec:todo}

Although the simultaneous fits have already been applied successfully to
real data (see Section~\ref{sec:gro1008} and \cite{kuehnel2013a}), the
routines and functions are still under development. We recommend to pull the
\texttt{isisscripts}-GIT-repository\footnotemark[2] regularly to be up-to-date.

There are, however, some caveats according to Table~\ref{tab:procontra} which
one should be aware of (as with any routine, not just our ISIS implementation).
In particular, the runtime still has to be reduced.
One way to achieve this is by performing the fit on multiple CPUs, e.g., one
CPU handles one dataset or datagroup. This has not been implemented yet because the
dependencies of the datasets on each other require data exchange between
the processes on the different machines. Additionally, the question about
weighing the data is currently under discussion. The weight depends on the
number of datapoints available in each dataset (or -group) as well as their
uncertainties - but what does this mean for its importance, i.e., its effect on
the model parameters? These remaining issues have to be clarified and the
respective solutions will be published in the future.

\begin{acknowledgements}
M.~K\"uhnel was supported by the Bundesministerium f\"ur Wirtschaft und
Technologie under Deutsches Zentrum f\"ur Luft- und Raumfahrt grant
50OR1113. We thank John E. Davis for developing the \texttt{SLXfig} package,
which was used to create all figures shown in this paper.
\end{acknowledgements}

\renewcommand{\bibsep}{0pt}

\end{document}